\documentclass[preprint,notoc]{JHEP3}
\usepackage{epsfig,cite,amsmath}

\def\to{\rightarrow}

\def\bi{\begin{itemize}}
 \def\ei{\end{itemize}}
\def\te{\tilde e}
\def\c1p{C1^\prime}

\def\tu{\tilde u}

\def\tb{\tilde b}

\def\tst{\tilde t}
\def\ttau{\tilde \tau}

\def\tg{\tilde g}
\def\tnu{\tilde\nu}

\def\tw{\tilde\chi}

\def\twpm{\tilde\chi^\pm}
\def\tz{\tilde\chi^0}
\def\alt{\stackrel{<}{\sim}}
\def\agt{\stackrel{>}{\sim}}
\def\be{\begin{equation}}  
\def\ee{\end{equation}}  
\def\bea{\begin{eqnarray}}  
\def\eea{\end{eqnarray}}

\def\Isajet{{\sc Isajet}}

\title{Is ``just-so'' Higgs splitting needed\\
for $t-b-\tau$ Yukawa unified SUSY GUTs?}
\author{Howard Baer$^{a}$, Sabine Kraml$^b$ and Sezen Sekmen$^c$\\
$^a$Dept.\ of Physics and Astronomy, University of Oklahoma, Norman, OK 73019, USA\\
$^b$Laboratoire de Physique Subatomique et de Cosmologie, UJF Grenoble 1, 
CNRS/IN2P3, INPG, 53 Avenue des Martyrs, F-38026 Grenoble, France\\
$^c$Dept.\ of Physics, Florida State University, Tallahassee, FL 32306\\
E-mail: \email{baer@nhn.ou.edu},\email{sabine.kraml@lpsc.in2p3.fr}, 
\email{sezen.sekmen@cern.ch}}

\preprint{\vbox{LPSC 09-109}}

\abstract{
Recent renormalization group calculations of the sparticle mass spectrum
in the Minimal Supersymmetric Standard Model (MSSM) 
show that $t-b-\tau$ Yukawa coupling unification at $M_{\rm GUT}$ 
is possible when the mass spectra follow the pattern of a
radiatively induced inverted scalar mass hierarchy. 
The calculation is entirely
consistent with expectations from $SO(10)$ SUSY GUT theories, 
with one exception: it seems to require MSSM Higgs soft term mass
splitting at $M_{\rm GUT}$, dubbed ``just-so Higgs splitting'' (HS) in the 
literature, which apparently violates the $SO(10)$ gauge symmetry.
Here, we investigate three alternative effects: 
{\it i}). $SO(10)$ $D$-term splitting, 
{\it ii}). inclusion of right hand neutrino in the RG calculation, and 
{\it iii}). first/third generation scalar mass splitting.
By combining all three effects (the DR3 model), we find 
$t-b-\tau$ Yukawa unification at $M_{\rm GUT}$ can be achieved at the 2.5\% level.
In the DR3 case, we expect lighter (and possibly detectable) 
third generation and heavy Higgs scalars than in the model with HS.
In addition, the light bottom squark in DR3 
should be dominantly a right state, while in the HS model, it is
dominantly  a left state.
}

\begin{document}

\section{Introduction}
\label{sec:intro}

The four LEP experiments performed precision measurements of the
$SU(3)_C$, $SU(2)_L$ and $U(1)_Y$ gauge coupling constants of the Standard
Model (SM) at energy scale $Q=M_Z$\cite{lep}. It is an astonishing fact that
the values of these couplings, evolved in energy from the weak scale to
an energy scale $M_{\rm GUT}\simeq 2\times 10^{16}$ GeV, nearly meet at a point
under Minimal Supersymmetric Standard Model (MSSM) renormalization
group (RG) evolution\cite{gaugeunif}, 
while their unification fails badly under Standard Model RG evolution.
This latter fact is often touted as indirect evidence that the MSSM 
(with weak-scale sparticle masses) is 
the correct effective field theory describing nature at energy scales
between $M_{\rm GUT}$ and $M_{\rm weak}$, and further that nature may well be 
described by a supersymmetric grand unified theory (SUSY GUT) 
at energy scales above $M_{\rm GUT}$.

While the gauge group $SU(5)$ early on emerged as a leading GUT group 
candidate\cite{su5}, 
the gauge group $SO(10)$ appears to be much more compelling\cite{so10}.
In SUSY $SO(10)$ GUTs, a number of attractive features emerge.
\begin{itemize}
\item All matter fields of a single generation are unified into the
16-dimensional spinor of $SO(10)$: thus, $SO(10)$ unifies matter as
well as forces.
\item The seemingly ad-hoc cancellation of triangle anomalies in the
SM and in $SU(5)$ is a simple mathematical fact in $SO(10)$.
\item The 16th element of the $SO(10)$ matter spinor is naturally occupied by
a superfield $\hat{N}^c$ which contains a SM-gauge singlet 
right-handed neutrino (RHN) field. 
Upon breaking of $SO(10)$, the RHN acquires a Majorana mass $M_N$ which leads
to the famous see-saw relation for neutrino masses\cite{seesaw}: 
$m_\nu =(f_\nu v_u )^2/M_N$.
\item The fact that matter superfields lie in a spinor representation
automatically leads to $R$-parity conservation, since only
superpotential terms of the form {\it matter-matter-Higgs} are allowed
by the $SO(10)$ symmetry, while the $R$-parity violating {\it matter-matter-matter} or
{\it matter-Higgs} products are not allowed. While $SO(10)$ breaking may
re-introduce $R$-parity violation, many simple breaking schemes 
exactly preserve the $R$-parity conserving structure.
\item $SO(10)$ SUSY GUTs naturally explain why two Higgs doublets
occur in nature at the weak scale. 
The ${\bf 2}$ and ${\bf 2^*}$ MSSM Higgs doublets lie in a 
${\bf 5}$ and ${\bf 5^*}$ of $SU(5)$, and the ${\bf 10}$ of $SO(10)$
naturally contains a ${\bf 5}$ and ${\bf 5^*}$ under restriction to
$SU(5)$.
\end{itemize}
Thus, $SO(10)$ SUSY GUTs provide an extremely compelling picture of 
what physics might look like around the GUT scale.

Along with gauge coupling and matter unification, the simplest
$SO(10)$ SUSY GUT models also predict third generation
$t-b-\tau$ Yukawa coupling unification at the GUT scale\cite{old}.
To check this assertion, one must begin with the measured third generation
fermion masses--- $m_t$, $m_b$ and $m_\tau$--- and calculate the associated
Yukawa couplings at the weak scale. Then one may evolve the $t-b-\tau$
Yukawa couplings up in energy to check whether they unify at $M_{\rm GUT}$, just as
the gauge couplings do. The values of the weak scale Yukawa couplings 
depend strongly on the ratio of Higgs field vevs: $\tan\beta\equiv v_u/v_d$. 
Furthermore, unlike the gauge couplings, the Yukawa couplings have a large
dependence on sparticle loop corrections when passing from the SM to the 
MSSM effective field theories\cite{hrs}. Thus, the $t-b-\tau$ Yukawa coupling
unification depends on the precise form of the sparticle mass spectra of
the MSSM. This latter fact offers a consistency check: if sparticle 
masses are found with the expected pattern, it would be strongly suggestive 
that an $SO(10)$ SUSY GUT model is valid around the GUT scale.

Detailed calculations of when $t-b-\tau$ Yukawa couplings unify
at $M_{\rm GUT}$ within the MSSM context have been performed by several
groups\cite{bf,bdr1,bdr2,abbbft,bkss,altm,bhkss}. 
They use an $SO(10)$-inspired model 
parameter space given by
\be
m_{16},\ m_{10},\ M_D^2,\ m_{1/2},\ A_0,\ \tan\beta, \ {\rm sign}(\mu )
\label{eq:pspace}
\ee
where $m_{16}$ is the GUT scale mass of all matter scalars, 
$m_{10}$ is the GUT scale mass of Higgs scalars, 
$M_D$ parametrizes Higgs mass splitting (HS) or possible scalar mass 
$D$-term splitting (DT)
(the latter can arise from the breaking of $SO(10)$ gauge symmetry), 
$m_{1/2}$ is the unified GUT scale gaugino mass, 
$A_0$ is the unified GUT scale soft SUSY breaking (SSB) trilinear term,
$\tan\beta\equiv v_u/v_d$ is the weak scale ratio of Higgs field vevs, 
and $\mu$ is the superpotential Higgs bilinear term, whose magnitude---
but not sign---is determined by the electroweak symmetry breaking (EWSB)
scalar potential minimization conditions. 

In practice, the two Higgs field soft breaking terms--- $m_{H_u}^2$ and 
$m_{H_d}^2$--- cannot be degenerate at $M_{\rm GUT}$ and still allow for an 
appropriate radiative breakdown of electroweak symmetry 
(REWSB)\cite{murayama}. 
Effectively, $m_{H_u}^2$ must be less than $m_{H_d}^2$ at $M_{\rm GUT}$ in order
to give $m_{H_u}^2$ a head start in running towards negative values 
at $M_{\rm weak}$. We parametrize 
the Higgs splitting as $m_{H_{u,d}}^2=m_{10}^2\mp 2M_D^2$ in accord with 
nomenclature for $D$-term splitting to scalar masses when a 
gauge symmetry undergoes a breaking which reduces the 
rank of the gauge group\cite{dterm}. 
While the $D$-term splitting should apply to matter scalar SSB terms as well,
in practice, better Yukawa unification for $\mu >0$ is found when the 
splitting is restricted only to the Higgs SSB terms. 
The mass splitting applied only to Higgs scalars, and not to matter scalars, 
has been dubbed ``just-so'' Higgs splitting in the literature\cite{bdr2}.

In previous work, the above parameter space was scanned over 
via random scans\cite{bf,abbbft} and also
by more efficient Markov Chain Monte Carlo (MCMC) scans\cite{bkss} to search 
for Yukawa unified solutions using the Isasugra subprogram 
of \Isajet\cite{isajet} for sparticle mass computations.\footnote{
Ref.~\cite{bhkss} confirms the general structure of Yukawa-unified models
also using the SoftSUSY\cite{softsusy} spectrum generator.}
The quantity
\be
R=\frac{max(f_t,f_b,f_\tau )}{min(f_t,f_b,f_\tau )}\ \ \ ({\rm evaluated\ at\ Q=M_{\rm GUT}}),
\label{eq:R}
\ee
was examined, where solutions with $R\simeq 1$ gave apparent Yukawa coupling 
unification.
For superpotential Higgs mass parameter $\mu >0$ (as favored by $(g-2)_\mu$ measurements),
Yukawa unified solutions with $R\sim 1$ were found
but only for {\it special} choices of GUT scale boundary conditions\cite{bf,bdr1,bdr2,abbbft,drrr,bkss,bhkss}:
\bi
\item  $m_{16}\sim 3-15~{\rm TeV}$,
\item  $A_0\sim -2m_{16}$,
\item  $m_{10}\sim 1.2 m_{16}$,
\item  $m_{1/2}\ll m_{16}$,
\item  $\tan\beta \sim 50$.
\ei
Models with this sort of boundary conditions were derived even earlier in 
the context of radiatively driven inverted scalar mass hierarchy models (RIMH) 
which attempt to reconcile suppression of
flavor-changing and $CP$-violating processes via a decoupling solution 
with naturalness 
via multi-TeV first/second generation and sub-TeV scale third generation scalars\cite{bfpz,bdqt}.
The Yukawa-unified spectral solutions were thus found in Refs.~\cite{abbbft,bkss} 
to occur with the 
above peculiar choice of boundary conditions as long as $m_{16}$ was in the {\it multi}-TeV regime.

Based on the above work\cite{abbbft,bkss}, the sparticle mass spectra from Yukawa-unified SUSY 
models are characterized qualitatively by the following conditions:
\begin{itemize}
\item first and second generation scalars have masses in the $\sim 10$ TeV regime,
\item third generation scalars, $\mu$ and $m_A$ have masses in the few TeV regime 
(owing to the inverted scalar mass hierarchy),
\item the gluino has mass $m_{\tg}\sim 300-500$ GeV,
\item the lightest neutralino $\tz_1$ is nearly pure bino with mass typically 
$m_{\tz_1}\sim 50-80$ GeV.
\end{itemize}

The presence of a bino-like $\tz_1$ along with multi-TeV scalars gives 
rise to a neutralino cold dark matter (CDM) relic abundance that 
is typically in the range $\Omega_{\tz_1}h^2\sim 10^2-10^4$, {\it i.e.} 
far above\cite{auto} the WMAP measured\cite{wmap5} value 
$\Omega_{CDM}h^2=0.110\pm 0.006$ by several orders of magnitude.
A very compelling solution to the Yukawa-unified dark matter abundance 
problem occurs if one invokes the Peccei-Quinn solution\cite{pq} 
to the strong $CP$ problem, 
which leads to dark matter being composed of an axion\cite{ww,axreview}/axino\cite{nillesraby,ckkr} 
admixture\cite{bkss,bs,bhkss}, instead of neutralinos.
The axino then can serve as the  LSP instead of the lightest neutralino\cite{wilczek,steff_rev}.
Cosmological solutions with a re-heat temperature $T_R$ high enough to sustain
non-thermal leptogenesis ($T_R\sim 10^6-10^8$ GeV) 
could most easily be found if the dominant component of the cold dark matter
consisted of {\it axions} rather than axinos.

The above calculational results show that Yukawa unified solutions are 
compatible with the MSSM as the effective theory between the GUT and 
weak mass scales, but for a very
constrained form of the sparticle mass spectrum. 
Indeed, the entire scheme is compatible with
expectations from an $SO(10)$ SUSY GUT theory with GUT symmetry broken at 
the scale $M_{\rm GUT}$, save for one feature. 
Naively, $SO(10)$ symmetry implies the two Higgs masses should be
{\it degenerate} at $M_{\rm GUT}$, at tree-level: 
{\it i.e.} $m_{H_u}^2=m_{H_d}^2$. 
The mechanism for the large GUT scale Higgs soft term splitting is unknown, 
and violates the $SO(10)$ symmetry. 
There is of course a well-known source of Higgs soft term splitting: namely, 
the $D$-term contribution to scalar masses which is induced when the rank 
of the gauge group is reduced from 5 of $SO(10)$ to 4 of $SU(5)$ or the SM. 
The $D$-term contributions are
necessarily proportional to the charge of the subgroup $U(1)_X$, and are to 
be included in {\it all} scalar fields carrying a $U(1)_X$ charge.
At $Q=M_{\rm GUT}$, the gauge symmmetry breaking induces
a scalar mass contribution
\begin{align}
m_Q^2=m_E^2=m_U^2 & = m_{16}^2+M_D^2 \label{eq:dt1} \\
m_D^2=m_L^2 & = m_{16}^2-3M_D^2 \label{eq:dt2} \\
m_{\tnu_R}^2 & = m_{16}^2+5M_D^2 \label{eq:dt3} \\
m_{H_{u,d}}^2 & =m_{10}^2\mp 2M_D^2, \label{eq:dt4}
\end{align}
where the contribution $M_D^2$ is effectively a free parameter of order the 
weak scale, and whose value depends on the details of $SO(10)$ breaking.
It can take positive or negative values.\footnote{Sum rules 
for sparticle masses as a test of the underlying $SO(10)$ are discussed in \cite{anant}.} 
It was found in Refs. \cite{bdr1,bdr2,abbbft} that calculationally, the spectral 
solutions with the best $t-b-\tau$ Yukawa unification occurred when the 
$D$-term splitting was applied only to the Higgs scalars, 
and not to the other matter scalars.

In this paper, we re-visit the question of the HS case versus the 
DT splitting case in $t-b-\tau$ Yukawa-unified models.
We search for Yukawa-unified solutions while including several effects 
which are all consistent with the general framework of simple 
$SO(10)$ SUSY GUT models:
{\it i}.) application of full DT splitting to all scalar masses, 
{\it ii}.) inclusion of neutrino Yukawa coupling effects (RHN)\cite{bdqt,bdr2}, and 
{\it iii}.) inclusion of mass splitting between the third generation, 
versus the first two generations of matter scalars (3GS).
We scan over $SO(10)$ model parameter space using the MCMC technique, 
which provides an efficient search for the best Yukawa unified solutions. 
We find that each of the above three effects
acts to improve the degree of Yukawa unification compared to results 
without the effects, but none of them work as well as the just-so HS model. 
However, using all three effects simultaneously (the DR3 model) 
does allow us to reach Yukawa-unified solutions with $R\sim 1.025$, 
{\it i.e.} Yukawa unification to below the 3\% level. 
The remaining last few percent might then be compatible with 
expected GUT scale threshold effects (which are of course model dependent)
and intrinsic theory error in our 2-loop RGE calculations. 
We find that the superparticle mass spectra using
the DR3 model is somewhat modified from solutions using just-so HS. 
In particular, for a given value of $m_{16}$, the
predicted value of $m_A$ and $m_{\tb_1}$ are much lighter than in 
the just-so HS prediction. 
For $m_{16}\sim 10$ TeV, $m_A$ and $m_{\tb_1}$ can extend down to or 
even below the 1 TeV level, and may be accessible to LHC searches 
and/or to searches at a future CERN $e^+e^-$ Linear Collider (CLIC),
where center-of-mass energies of order $\sqrt{s}\sim 3-5$ TeV are proposed. 
The spectral differences, and the expected left-right composition of the 
lightest bottom squark, 
may offer a means to distinguish between the just-so HS model
and the DR3 model.

\section{Higgs mass splitting and radiative EWSB in Yukawa unified models}
\label{sec:rewsb}

\subsection{Radiative EWSB in Yukawa unified models}

MSSM models with $t-b-\tau$ Yukawa coupling unification
and degenerate (Higgs) soft masses at $Q=M_{\rm GUT}$ 
face a well-known problem when one attempts to generate
realistic sparticle mass spectra: electroweak symmetry
fails to be appropriately broken\cite{murayama,old}. The problem 
can be seen by examining the one-loop Higgs soft mass
RGEs:
\bea
\frac{dm_{H_d}^2}{dt}&=&{2\over 16\pi^2}\left(-{3\over 5}g_1^2M_1^2-
3g_2^2M_2^2-{3\over 10}g_1^2S+3f_b^2X_b+f_\tau^2X_\tau\right)\,, 
\label{eq:mhd} \\
\frac{dm_{H_u}^2}{dt}&=&{2\over 16\pi^2}\left(-{3\over 5}g_1^2M_1^2-
3g_2^2M_2^2+{3\over 10}g_1^2S+3f_t^2X_t\right)\,,\label{eq:mhu}
\eea
where 
\bea
X_t&=&m_{Q_3}^2+m_{\tst_R}^2+m_{H_u}^2+A_t^2\, ,\label{eq:xt} \\
X_b&=&m_{Q_3}^2+m_{\tb_R}^2+m_{H_d}^2+A_b^2\, ,\label{eq:xb} \\
X_\tau &=&m_{L_3}^2+m_{\ttau_R}^2+m_{H_d}^2+A_\tau^2\,, \label{eq:xl} 
\eea
and 
\be
S=m_{H_u}^2-m_{H_d}^2+Tr\left[{\bf m}_Q^2-{\bf m}_L^2-2{\bf m}_U^2
+{\bf m}_D^2+{\bf m}_E^2\right] .\label{eq:S}
\ee
The right-hand side terms with negative co-efficients
give an {\it upwards} push to the Higgs soft masses during evolution
from $M_{\rm GUT}$ to $M_{\rm weak}$, while the positive terms give a
{\it downwards} push. 

At the weak scale, where the Higgs effective potential is minimized,
the EWSB minimization conditions require that (at tree-level)
\bea
B\mu &=& \frac{(m_{H_u}^2+m_{H_d}^2+2\mu^2 )\sin 2\beta}{2}\,, \label{eq:B} \\
\mu^2 &=& \frac{m_{H_d}^2-m_{H_u}^2\tan^2\beta}{(\tan^2\beta -1)}
-\frac{M_Z^2}{2}\, . \label{eq:mu}
\eea
The first of these determines the weak scale value of $B$
in terms of $\tan\beta$; the second relation determines the
magnitude, but not the sign, of the superpotential $\mu$ term.
At moderate-to-large $\tan\beta$ values and $|m_{H_u}^2|\gg M_Z^2$, 
the second relation also gives approximately $\mu^2\simeq -m_{H_u}^2$,
and we see that $m_{H_u}^2$ must be driven to negative values
to accommodate successful REWSB.
In models with Yukawa couplings $f_t> f_b,\ f_\tau$, the
$m_{H_u}^2$ term is pushed to negative values by the large
value of $f_t$ in Eq.~(\ref{eq:mhu}), resulting in successful EWSB. 
In contrast, in models with $t-b-\tau$ Yukawa unification,
the Yukawa coupling terms on the right-hand side of the 
$m_{H_d}^2$ equation are {\it larger} than the 
corresponding terms in the $m_{H_u}^2$ equation, resulting
in $m_{H_d}^2$ being driven more negative than $m_{H_u}^2$ at
the weak scale. If $m_{H_d}^2< m_{H_u}^2\tan^2\beta$, then 
$\mu^2<0$, signaling an inappropriate EWSB.  
The solution to this dilemma so far in Yukawa unified models is
to provide the $m_{H_u}^2$ term a {\it head-start} in running to negative
values at the weak scale by adopting Higgs splitting such that 
$m_{H_u}^2< m_{H_d}^2$ at the GUT scale. 

While just-so HS applies a splitting
(that violates $SO(10)$ gauge symmetry) only to Higgs soft masses,
and leaves the remaining GUT scale scalar masses fixed at $m_{16}$,
the expected splitting due to $D$-terms arising from $SO(10)$ 
breaking at $M_{\rm GUT}$ apply to matter scalars as well as Higgs scalars
as given in Eqs.~(\ref{eq:dt1})--(\ref{eq:dt4}).
The problem arising from DT splitting is that the $m_D^2$ terms
are substantially reduced already at $M_{\rm GUT}$, and can get driven tachyonic
at $M_{\rm weak}$ through the RIMH mechanism.
What is needed in terms of DT splitting is a large enough
Higgs splitting to facilitate REWSB, but not so large a splitting that
$m_{\tb_R}^2$ is driven tachyonic at $M_{\rm weak}$.

\subsection{The role of neutrino Yukawa couplings}

An alternative mechanism to aide $m_{H_u}^2$ being driven negative
is to {\it balance} the right-hand-side Yukawa push in Eqs.~(\ref{eq:mhd})--(\ref{eq:mhu})
by incorporating the effect of the third generation neutrino
Yukawa coupling. In this case, the $m_{H_u}^2$ soft term RGE is modified 
to
\bea
\frac{dm_{H_u}^2}{dt}=\frac{2}{16\pi^2}\left[-{3\over 5}g_1^2M_1^2-
3g_2^2M_2^2+3f_t^2X_t+f_\nu^2X_\nu \right] \label{eq:mhu_rhn}
\eea 
with $X_\nu =m_L^2+m_{\tnu_R}^2+m_{H_u}^2+A_\nu^2$, and the RGE for
$m_{H_d}^2$ is unchanged.
We see that for unified Yukawa couplings, now $m_{H_u}^2$ receives
an {\it additional} downward push from the $f_\nu$ term\cite{bdqt,bdr2}. The $f_\nu$ term
contributes to the running at energy scales $M_N<Q<M_{\rm GUT}$. For $Q<M_N$, 
the RHN states are integrated out of the effective theory, and the 
RGEs revert to the MSSM form of Eq.~(\ref{eq:mhu}).

\subsection{Third generation splitting}

A third effect that can improve the implementation of EWSB in Yukawa unified
models is to allow for non-degeneracy in first/second versus third
generation scalar masses: third generation mass splitting (3GS). 
In general, generational non-degeneracy can lead
to flavor violation in excess of experimental bounds\cite{masiero}. 
The bounds on flavor-changing neutral currents (FCNC) apply most 
strongly to splitting amongst the first and second generation 
soft terms; contraints on third generation FCNCs are much less 
restrictive\cite{nmh}. Here, we will maintain degeneracy between $m_{16}(1)$
and $m_{16}(2)$, although some small breaking in these terms
is allowed, especially in our case where scalar masses are quite heavy and 
typically in the 10 TeV range, where we also have FCNC 
suppression via decoupling. 
Here, we will adopt $m_{16}(1)=m_{16}(2)\ne m_{16}(3)$. 

At tree level, we would expect 3GS to not affect Yukawa coupling
evolution and EWSB, since the Higgs sector only couples strongly to third
generation scalars, and is independent of first/second generation
scalar masses. However, at two-loop level, the scalar mass RGEs have the form
given by\cite{mv}
\be
\frac{dm_i^2}{dt}=\frac{1}{16\pi^2}\beta_{m_i^2}^{(1)}+
\frac{1}{(16\pi^2)^2}\beta_{m_i^2}^{(2)},
\ee
where $t=\ln Q$,
$i=Q_j,\ U_j,\ D_j,\ L_j$ and $E_j$, and $j=1-3$ is a generation index.
Two loop terms are formally suppressed relative to one loop terms by
the square of a coupling constant as well as an additional loop
factor of $16\pi^2$. However, these two loop terms include contributions
from {\it all} scalars. Specifically, the two loop $\beta$ functions include,
\be
\beta_{m_i^2}^{(2)}\ni a_ig_3^2\sigma_3+b_ig_2^2\sigma_2+c_ig_1^2\sigma_1,
\ee
where
\bea
\sigma_1 &=& {1\over 5}g_1^2\{3(m_{H_u}^2+m_{H_d}^2)+Tr[{\bf m}_Q^2+
3{\bf m}_L^2+8{\bf m}_U^2+2{\bf m}_D^2+6{\bf m}_E^2]\},\nonumber \\
\sigma_2 &=& g_2^2\{m_{H_u}^2+m_{H_d}^2+Tr[3{\bf m}_Q^2+
{\bf m}_L^2]\},\ \ \ \ {\rm and} \nonumber\\
\sigma_3 &=& g_3^2Tr[2{\bf m}_Q^2+{\bf m}_U^2+{\bf m}_D^2],\nonumber
\eea
and the ${\bf m}_i^2$ are squared mass matrices in generation space.
The numerical coefficients $a_i$, $b_i$ and $c_i$ are related to the quantum
numbers of the scalar fields, but are all positive quantities. 
Incorporation of large, multi-TeV masses for the first and second generation
scalars leads to an overall positive, {\it non-negligible}
contribution to the slope of SSB mass trajectories versus energy 
scale\cite{gsimh}.
Although formally a two loop effect,
the smallness of the couplings is compensated by the
multi-TeV scale values of masses for the first two generations of scalars.
In running from $M_{\rm GUT}$ to $M_{\rm weak}$,
the two-loop terms result in an overall {\it reduction} of scalar masses, 
and its effect depends on the quantum numbers of the
various scalar fields.

Generational non-degeneracy of scalar masses, especially for
the 3GS scenario, is natural in $SO(10)$ SUSY GUT models where
above-the-GUT-scale-running is allowed. In this case, the unified
third generation Yukawa coupling acts to suppress $m_{16}(3)$
with respect to $m_{16}(1,2)$, even if the three generations are degenerate
at some higher scale, {\it e.g.} at the Planck scale $M_P$.

$SO(10)$ model RGEs are presented in Sec. 6.1 of Ref.~\cite{bdqt}.
As an example, we show in Fig.~\ref{fig:rge} the evolution of
$m_{10}^2$, $m_{16}(1,2)$ and $m_{16}(3)$ from $M_P$ to $M_{\rm GUT}$ 
for model parameters as depicted in the caption. We see that a splitting of order
25\% is possible at $M_{\rm GUT}$ for parameter choices as befitting Yukawa unified
models. The natural splitting here is then that $m_{16}(3)< m_{16}(1,2)$, 
assuming degeneracy at $M_{P}$. In our following calculations, 
we will merely implement $m_{16}(3)$ as a free parameter as 
distinct from $m_{16}(1,2)$.

\FIGURE[t]{
\includegraphics[width=9cm]{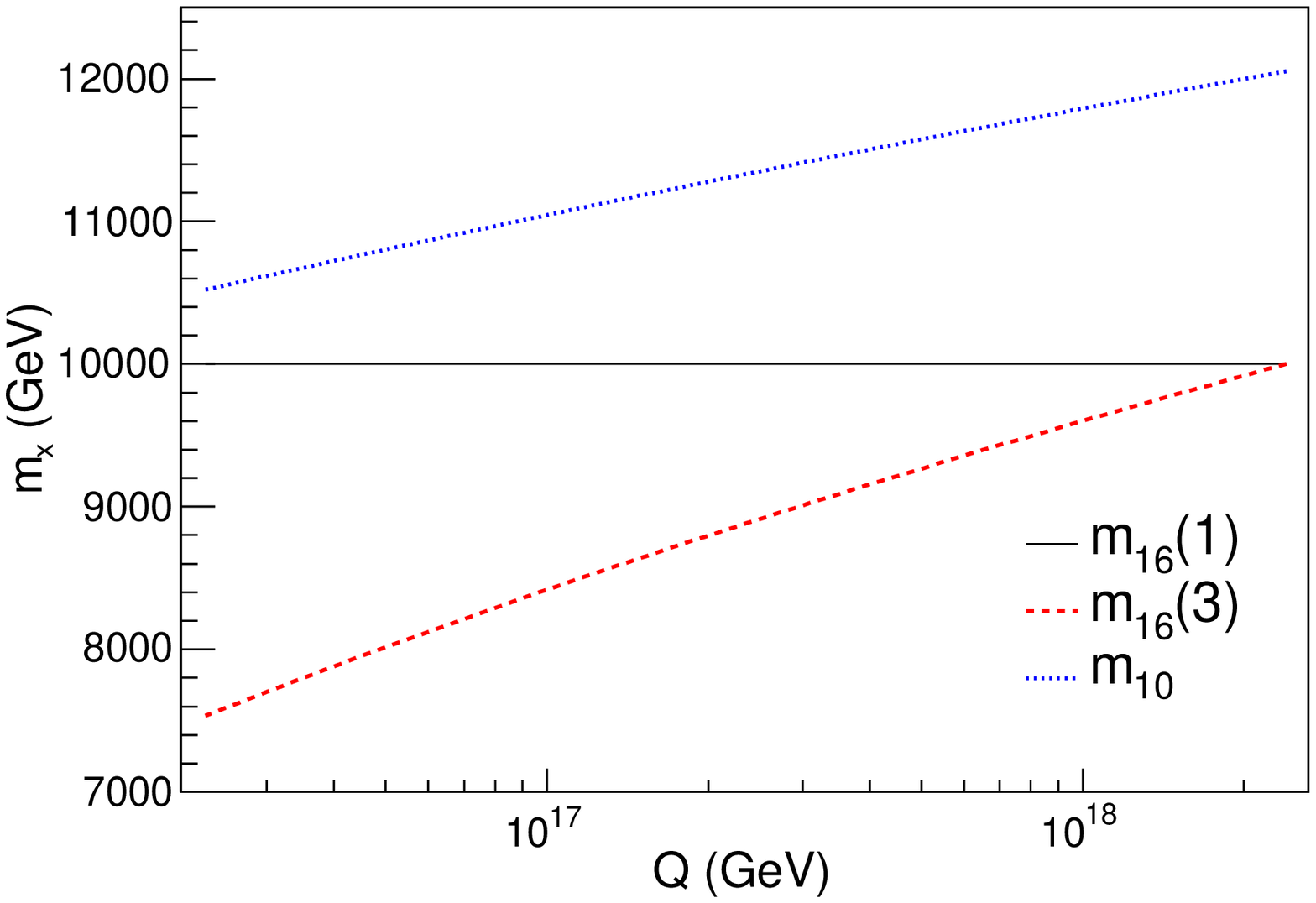}
\caption{Renormalization group evolution of soft SUSY breaking
terms $m_{16}(1,2)$, $m_{16}(3)$ and $m_{10}$ from 
$M_P$ to $M_{\rm GUT}$ in a simple $SO(10)$ SUSY GUT. 
We take $g_{GUT}=0.682$, $f_{GUT}=0.56$, $m_{16}=10$ TeV, 
$m_{10}=12.05$ TeV and $A_0=-19.947$ TeV (point~B of Table~2
of Ref.~\cite{bhkss}).
}\label{fig:rge}}

\section{Numerical results for Yukawa unified models}
\label{sec:results}
\subsection{MCMC scan of parameter space}

In this section, in addition to the just-so HS model, 
we will also scan over the $SO(10)$ model parameter space as given by
\be
m_{16}(1,2),\ m_{16}(3),\ m_{10},\ M_D,\ m_{1/2},\ A_0,\ \tan\beta ,\ sign(\mu ), \label{eq:pspace2}
\ee
where we also update our boundary conditions of  $m_b(M_Z)^{\overline{DR}}=2.416$,
$\alpha_s (M_Z)=0.118$ and
$m_t=173.1$ GeV; the latter in accord with recent Tevatron 
measurements\cite{mtop}. In addition,
we augment the parameter space with the RHN parameter set
\be
M_{N_3},\ f_{\nu_\tau},\ A_{\nu_\tau},\ m_{\tnu_{R3}}
\ee
where we assume $f_{\nu}=f_t$ at $Q=M_{\rm GUT}$, $A_\nu =A_0$ and
$m_{\tnu_{R3}}$ is as given in Eq.~(\ref{eq:dt3}) by DT splitting.
For $M_{N_3}$, we are guided by the simple see-saw relation
\be
m_{\nu_3}\simeq \frac{(f_{\nu_\tau}v_u)^2}{M_{N_3}}
\ee
where for $\tan\beta\sim 50$, we have $v_u(m_{\rm SUSY})\sim 171.6$ GeV,
$v_d(m_{\rm SUSY})\sim 3.5$ GeV and $f_{\nu_\tau}\sim 0.54$. Then,
cosmological bounds on the sum of neutrino masses implies
$\sum m_{\nu_i}\alt 1$ eV. This implies $M_{N_e}\agt 10^{13}$ GeV.
We will adopt here $M_{N_3}=10^{13}$ GeV to maximize the downward
push of $f_{\nu_\tau}$ consistent with bounds on neutrino masses.
The superparticle mass spectrum is then generated using 
Isajet 7.79\cite{isajet}, which includes
full two-loop RGE running, implementation of the RG-improved
1-loop effective potential for EWSB, and full 1-loop corrections
to all sparticle masses.
We scan over the parameter space using the MCMC method with a 
Metropolis sampling algorithm as in \cite{bkss}, 
which provides an optimized search for parameter
space points with the lowest $R$ values. 

In the scans, we require that the mass limits from direct SUSY \cite{lepsusy} and 
Higgs \cite{lephiggs} searches at LEP be observed (additional limits from Tevatron 
searches do not affect the solutions with small $R$). 
Moreover, we take into account the constraints from the branching fractions 
for the $b\to s\gamma $ and $B_s\to\mu^+\mu^-$ decays. 
The measured branching ratio of the inclusive radiative $B$ decay is 
BR$(b\to s\gamma )=(3.52\pm 0.23\pm0.09)\times 10^{-4}$ \cite{Barberio:2008fa}, 
and the SM theoretical prediction  
BR$(b\to s\gamma )^{\rm SM}=(3.15\pm 0.23)\times 10^{-4}$ \cite{Misiak:2006zs}. 
Combining experimental and theoretical errors in quadrature, we take  
$2.85\le {\rm BR}(b\to s\gamma)\times 10^4 \le 4.19$ at $2\sigma$ 
together with the 95\% CL upper limit BR$(B_s\to\mu^+\mu^-)<5.8\times 10^{-8}$ from 
CDF~\cite{Aaltonen:2007kv}.  
We adopt the \Isajet\ Isatools\cite{isatools} program for the calculations 
of BR$(b\to s\gamma)$ and BR$(B_s\to\mu^+\mu^-)$.

Our results for parameter space scans with minimized $R$ values
are summarized in Table \ref{tab:R}. 
We first see that $t-b-\tau$ Yukawa unification is not possible
in the mSUGRA model: in this case,  $R_{min}=1.35$. 
However, adopting $SO(10)$ parameter space
with RHN parameters (but with $M_D=0$) actually allows considerable
improvement with $R$ values as low as 1.19 to be attained. 
The same $R_{min}$ can be obtained in mSUGRA+RHN.  
Allowing $SO(10)$+DT splitting (but no RHN), we find $R_{min}=1.08$, while
$SO(10)$+HS yields $R_{min}=1.0$, as noted in many previous studies.
An allowance of DT+3GS yields a value $R_{min}=1.06$
while allowing DT+RHN gives $R_{min}=1.04$.
$SO(10)$ scans with DT+RHN+3GS (the DR3 model) allow 
us to generate models with $R_{min}$ values down to 1.025. The remaining few 
per cent may be accounted for either by GUT scale threshold corrections,
or theoretical error due to our imperfect modeling of Yukawa coupling
boundary conditions and evolution over $13-14$ orders of magnitude.
If these combined effects are at the level of few per cent, 
then the $SO(10)$ model with D-term splitting plus RHN and/or 3GS 
can be seen to be in accord with Yukawa unified models for $\mu >0$.
In the following, we concentrate on the DR3 model, which combines 
DT+RHN+3GS and gives the best prospects for Yukawa unification, 
and compare it to the ``just-so'' Higgs splitting. 

\begin{table}\centering
\begin{tabular}{lc}
\hline
model & $R_{min}$ \\
\hline
mSUGRA       & 1.35 \\
mSUGRA+RHN       & 1.19 \\
$SO(10)$+RHN, $M_D=0$       & 1.19 \\
$SO(10)$+DT    & 1.08 \\ 
$SO(10)$+3GS & 1.30 \\
$SO(10)$+DT+3GS    & 1.06 \\ 
$SO(10)$+DT+RHN    & 1.04 \\ 
$SO(10)$+RHN+3GS    & 1.17 \\ 
$SO(10)$+DT+RHN+3GS (DR3)    & 1.025 \\ 
$SO(10)$+HS    & 1.0 \\ 
\hline
\end{tabular}
\caption{Minimal $R$ values obtained from MCMC 
scans for $t-b-\tau$ Yukawa unification in
various model parameter space choices.
}
\label{tab:R}
\end{table}

Next, we examine which parameter choices lead to $t-b-\tau$ Yukawa
coupling unification in the DR3 model.
To this end, we plot the locus of Yukawa unified solutions
with $R<1.05$ in Fig.~\ref{fig:mDvm16_3} in the $m_{16}(3)/m_{16}\ vs.\ M_D$
plane.\footnote{
Note that in our case $M_D^2$ is always positive since we need $m_{H_u}^2<m_{H_d}^2$ 
at the GUT scale.} 
Here and in the following, $m_{16}\equiv m_{16}(1,2)$ for simplicity. 
The red points indicate results from the just-so HS model.
They necessarily all have $m_{16}(3)/m_{16}=1$, and so form a 
vertical line through the plot. 
The values of $M_D$ in the HS model actually range up to $\sim 5$ TeV.
In contrast, we find that the DR3 model yields
Yukawa-unified solutions provided that $m_{16}(3)\sim (0.8-1.05)m_{16}(1,2)$:
{\it i.e.} the third generation scalars are favored to be at 
somewhat lower GUT scale masses than their first/second generation
counterparts. In addition, we see that $M_D$ is now restricted to lie
in the $1-2$ TeV range: significantly less than the 
splitting needed for the HS model.

\FIGURE[t]{
\includegraphics[width=9cm]{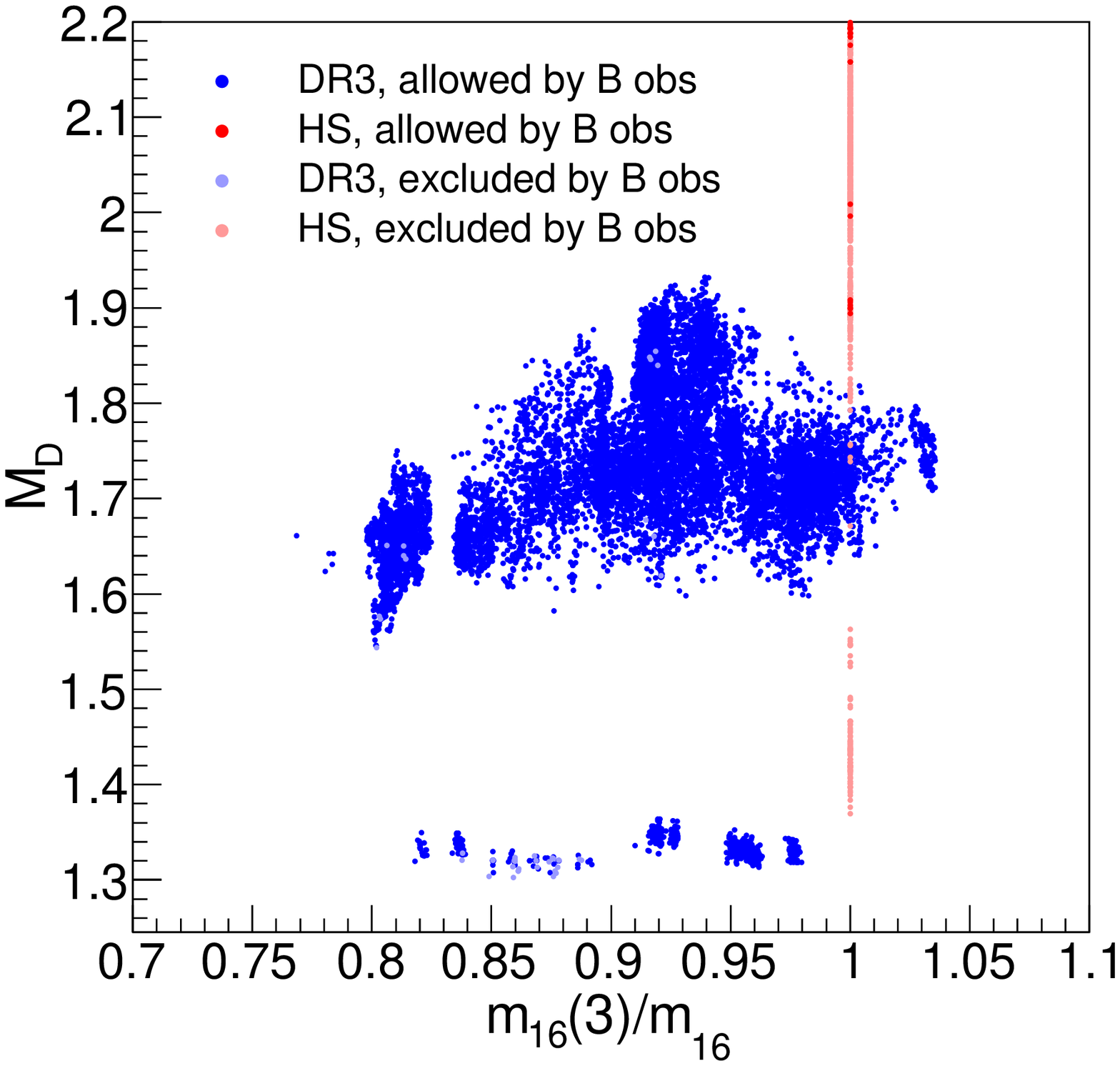}
\caption{Yukawa unified solutions with $R<1.05$ from ``just-so'' HS model (red points) 
and the DR3 model (blue points) in the $m_{16}(3)/m_{16}\ vs.\ M_D$ plane. 
Light blue/light red points are excluded by $B$-physics constraints. 
Note that $M_D$ extends up to about 5~TeV in the HS case, cf.\ Fig.~\ref{fig:mDvm16}.
}\label{fig:mDvm16_3}}

In Fig.~\ref{fig:mDvm16}, we show the locus of points with
$R<1.05$ in the $M_D\ vs.\ m_{16}$ plane. Here we see that for
the HS model, values of $M_D\sim 0.33 m_{16}$ are needed. In the
case of the DR3 model, we see that the value of 
$M_D$ needed also grows with $m_{16}$. But in this case, we find
instead that $M_D\sim 0.13 m_{16}$. The amount of Higgs splitting
needed in the DR3 model is much less than in the
HS model since additional Higgs splitting comes from 
the effect of the RHN and 3GS.
Other than the relative reduction in $M_D$ needed for a given value
of $m_{16}(1,2)$, the usual SSB mass relations for Yukawa-unified HS model
still hold. 

\FIGURE[t]{
\includegraphics[width=9cm]{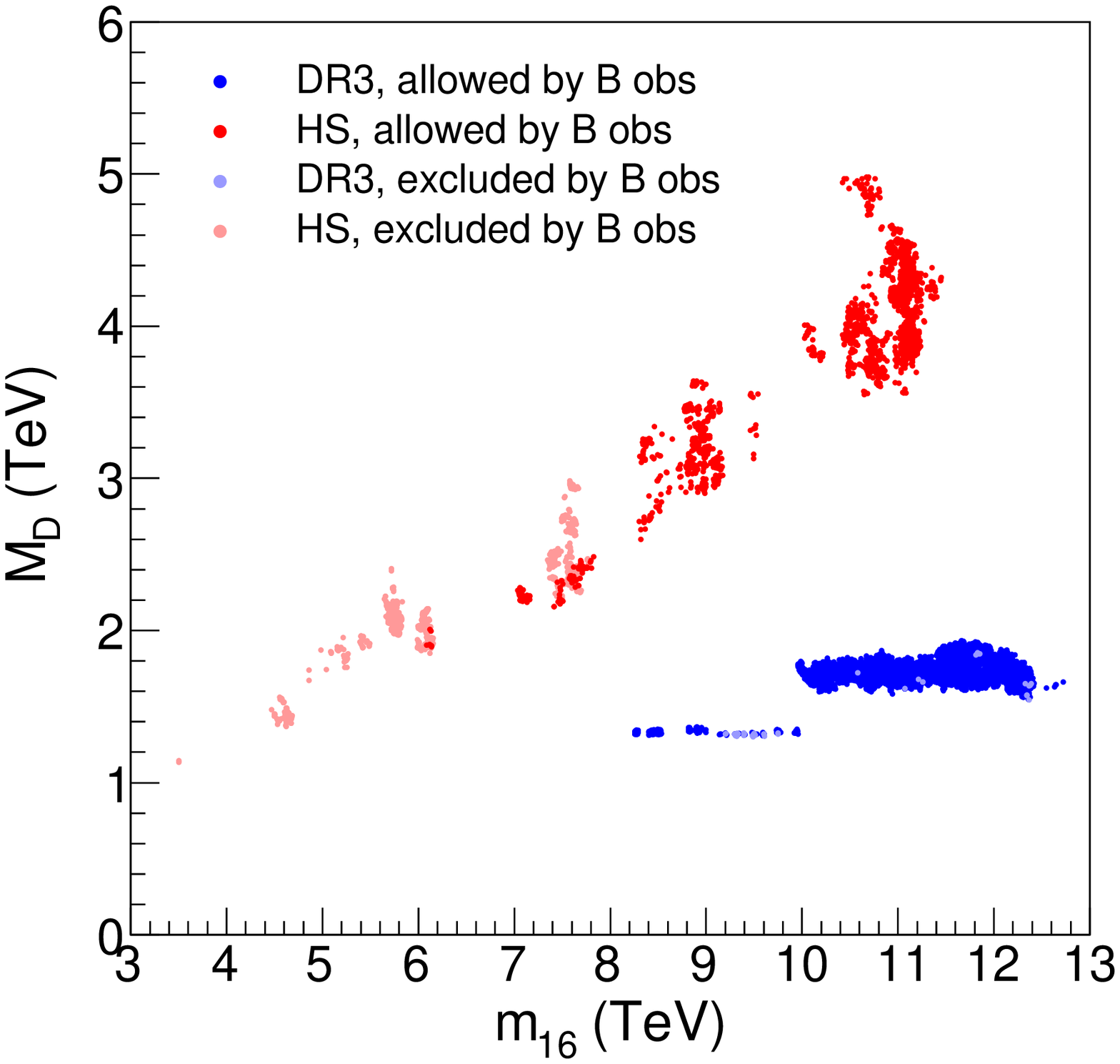}
\caption{Yukawa unified solutions with $R<1.05$
from ``just-so'' HS model (red points) and models with
DR3 splitting (blue points) in the $M_D\ vs.\ m_{16}$ plane; 
points excluded by $B$-physics constraints are shown in lighter colour.
}\label{fig:mDvm16}}

Note also that in the DR3 model, Yukawa-unified solutions occur only for 
$m_{16}(1,2)\agt 8$~TeV, while in the HS model they also occur for smaller 
values of $m_{16}$. However, the HS points with 
$m_{16}\sim 4-8$~TeV are almost all excluded by BR($b\to s\gamma $) 
and/or BR($B_s\to\mu^+\mu^-$). This observation is also made by Altmannshofer 
et al.\cite{altm}. The DR3 model points, on the other hand, are much less affected 
by the $B$-physics constraints. We will come back to this in section~3.3.

\subsection{SUSY particle mass spectrum}

Given that both the HS and DR3 models
lead to $t-b-\tau$ unification at $M_{\rm GUT}$, the next question is
whether it is possible to physically distinguish between these
models at experiments. 
Several differences between the SUSY particle mass spectrum lead us to
believe that the models are at least in principle distinguishable. 
The first point is that--- in the case of the DR3 model--- the
GUT scale soft masses for $m_D^2$ and $m_L^2$ are diminished by 
$-3M_D^2$ relative to the value of $m_{16}$. The first and second generation
values of $m_D$ and $m_L$ are expected to be in the multi-TeV regime, and so
their mass diminution by $D$-terms isn't likely to be visible at any
collider expected to operate in the near future. 
However, the third generation scalar
masses are driven to weak scale values by the RIMH mechanism, and are
expected to be in the 1--2 TeV regime. Thus, we would expect the
third generation $\tb_R$ and $\ttau_L$ masses to be diminished with respect to
expectations from the HS model. This effect should be most noticable in the
bottom squark sector, since in the tau slepton sector, we usually expect
(based on the form of the MSSM RGEs) $m_{\ttau_L}> m_{\ttau_R}$, 
whereas in the sbottom sector, we expect $m_{\tb_R}< m_{\tb_L}$.

In Fig.~\ref{fig:mt1}, we show the value of $m_{\tst_1}\ vs.\ m_{16}$ for 
Yukawa-unified models with $R<1.05$ in the HS (red dots) and DR3 (blue dots) cases. 
Points which obey the mass limits but do not pass $B$-physics constraints 
are again shown in lighter colour. We see that for a given
value of $m_{16}$, the value of $m_{\tst_1}$ is smaller in the DR3 case than in the 
HS case. Naively, one might expect the opposite result, since in the DR3 model
both $m_{\tst_L}^2$ and $m_{\tst_R}^2$ are {\it increased} by $+M_D^2$.
However, two effects act counter to the $D$-term. First, there is the third generation
mass splitting, which typically reduces $m_{16}(3)$ relative to $m_{16}(1)$. 
Second, there is an additional contribution to RG running of top squark soft masses 
from the $S$-term, Eq.~(\ref{eq:S}), which is zero for models with strict universality, but 
which is non-zero for models with Higgs mass splitting. For our case with 
$m_{H_u}^2<m_{H_d}^2$, 
the $S$-term gives an {\it upwards} push to the top squark soft mass evolution.
Since the Higgs splitting is much less in the DR3 model, there is a smaller
upwards push from the $S$-term, and this effect coupled with 3GS wins out 
over the increased mass due to the $D$-term, 
thus giving the DR3 models typically a smaller $\tst_1$ mass than in the HS case. 
We note here that while the value of $m_{\tst_1}$ is smaller in the DR3 case compared to HS-- for
a given value of $m_{16}$-- the value of $m_{16}$ will not be easily measureable, and so 
the $m_{\tst_1}/m_{16}$ ratio is not likely a good discriminator between the two models:
for instance, if a value of $m_{\tst_1}\sim 1.5$ TeV is found at some future
experiment, it will be difficult to
determine if it is consistent with the HS model with $m_{16}\sim 7$ TeV, or with the DR3 model with
$m_{16}\sim 9$ TeV.
\FIGURE[t]{
\includegraphics[width=9cm]{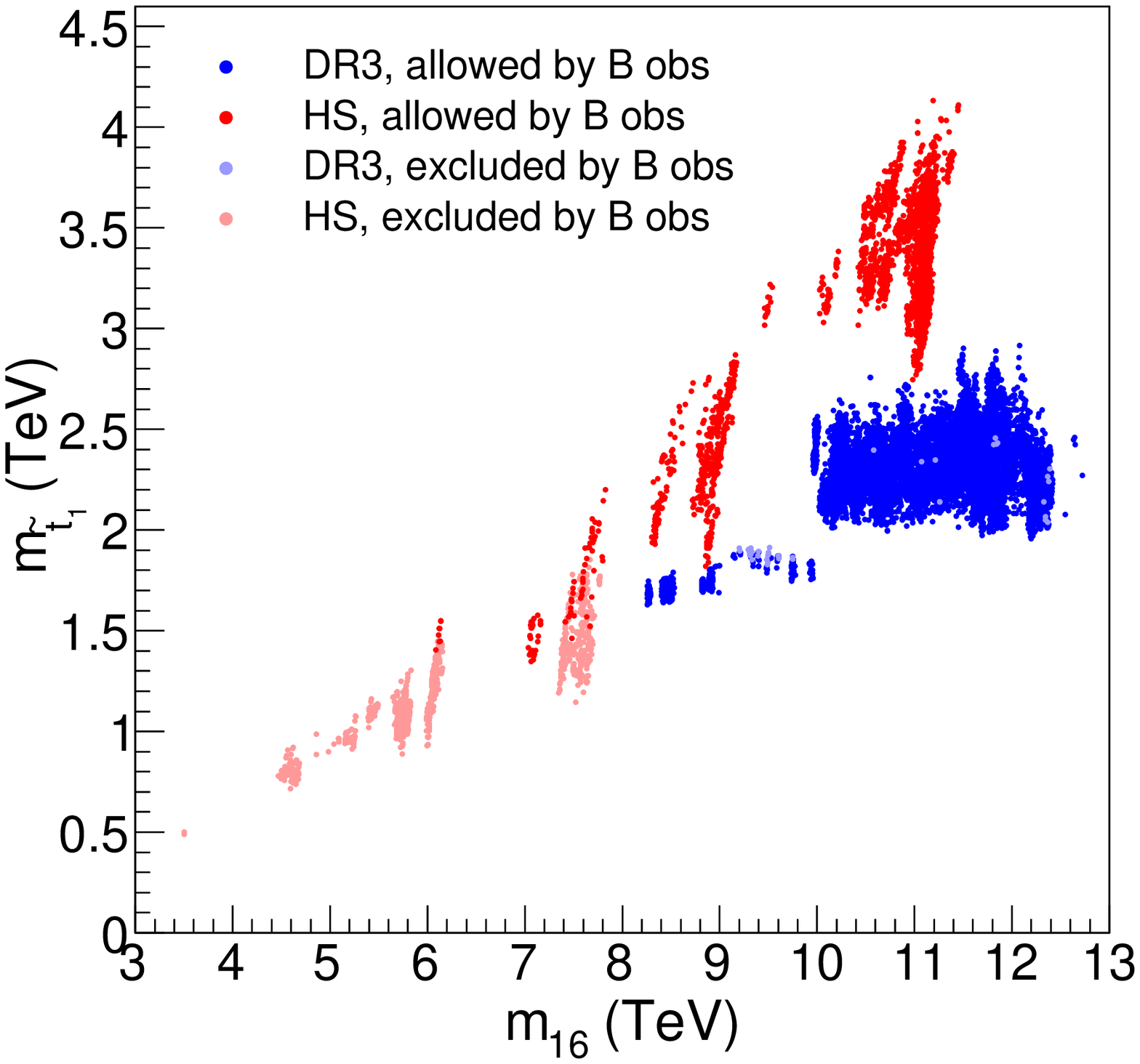}
\caption{Yukawa unified solutions with $R<1.05$ from ``just-so'' 
HS model (red points) and models with DR3 splitting (blue points) 
in the $m_{16}\ vs.\ m_{\tst_1}$plane; points excluded by $B$-physics 
constraints are shown in lighter colour.
}\label{fig:mt1}}

In Fig.~\ref{fig:mb1}, we plot the value of $m_{\tb_1}$ expected from 
Yukawa-unified models with HS (red dots) and DR3 (blue dots), 
where again we only plot solutions with $R<1.05$. We see that in the HS case, 
for a given value of $m_{16}$, the value of $m_{\tb_1}$ is always
lowest in the DR3 case. In fact, for $m_{16}\sim 10$ TeV, we expect 
$m_{\tb_1}\sim 1-2$ TeV, while in the HS case, $m_{\tb_1}\sim 3-4$ TeV. 

\FIGURE[t]{
\includegraphics[width=9cm]{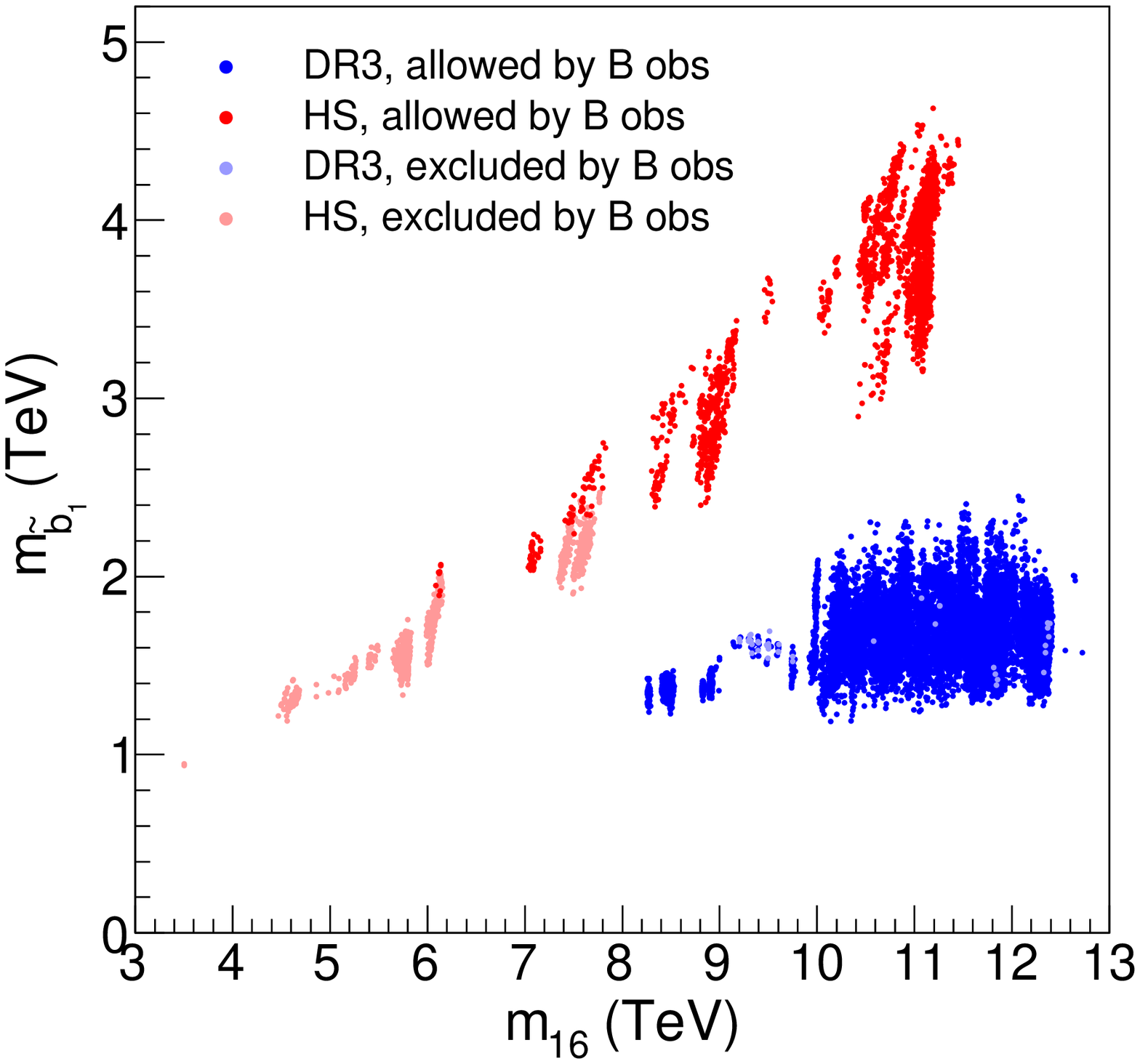}
\caption{Yukawa unified solutions with $R<1.05$ from the ``just-so'' 
HS model (red points) and models with DR3 splitting (blue points) 
in the $m_{16}\ vs.\ m_{\tb_1}$ plane; points excluded by $B$-physics 
constraints are shown in lighter colour.
}\label{fig:mb1}}

The mixing angle of $\tb_1$ is also strongly affected. Here, following the
notation of Ref.~\cite{wss}, we have $\tb_1 =\cos\theta_b \tb_L -\sin\theta_b\tb_R$.
As an illustrative example, we list in Table \ref{tab:mass} the spectrum from a HS 
Yukawa-unified point with $m_{16}=10$ TeV, and a DR3 model case with $m_{16}=11.8$~TeV. 
The evolution of all four Yukawa coupings are shown for the  DR3 point
in Fig.~\ref{fig:yuk}.

\FIGURE[t]{
\includegraphics[width=9cm]{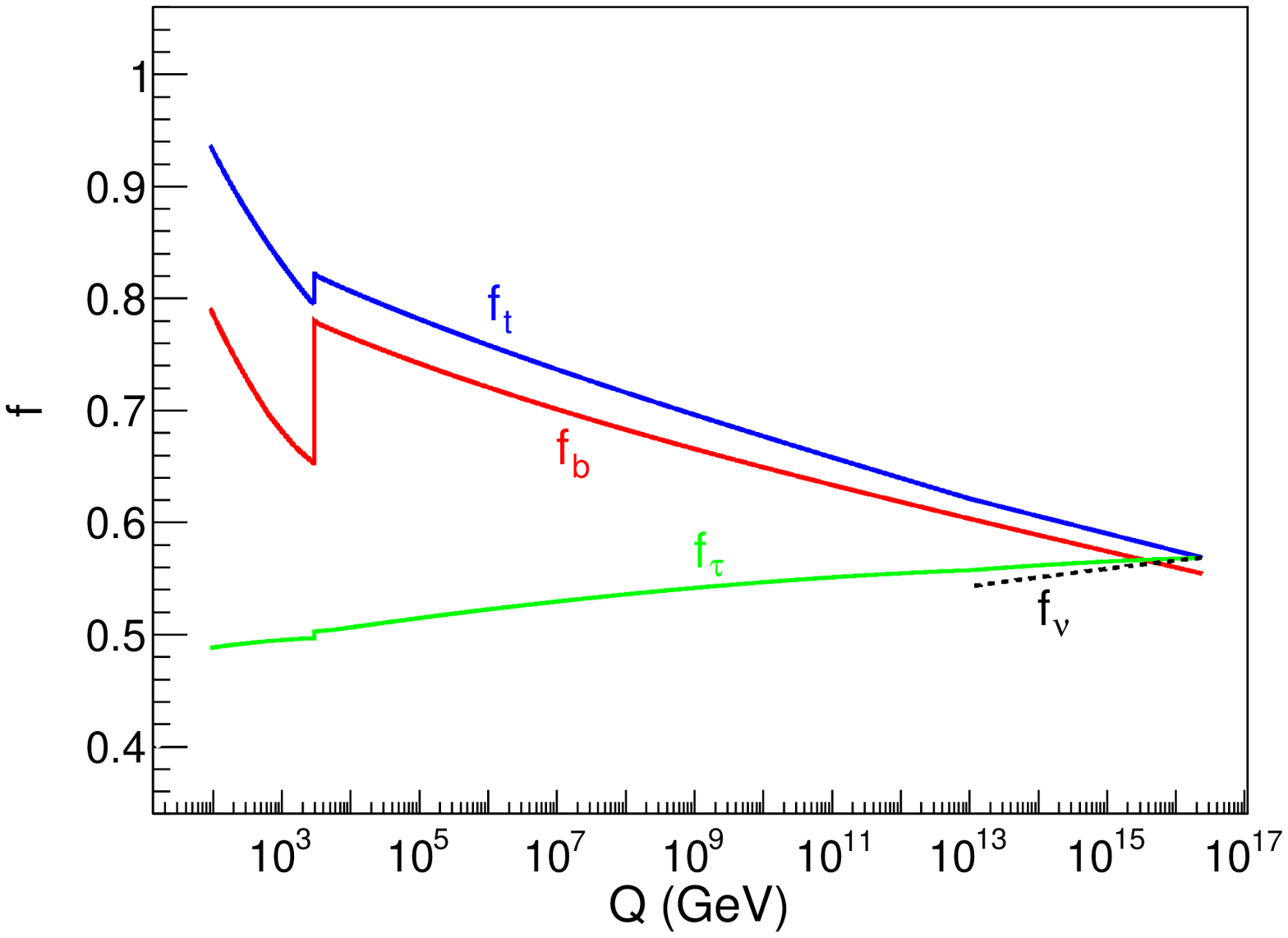}
\caption{Evolution of all four Yukawa couplings for
the case of point DR3 listed in Table \ref{tab:mass}.
}\label{fig:yuk}}

In the HS case, $\tb_1$ is $\sim 10\%$ $\tb_R$, 
while in the DR3 case, $\tb_1$ is $99.8\%$ $\tb_R$. 
This comes from the fact that the $D$-term mass contribution pushes $m_{\tb_L}$ up, 
and $m_{\tb_R}$ down. We also show in Fig.~\ref{fig:thb} the $b$-squark mixing
angle $\theta_b$ versus $m_{\tb_1}$. In this plot, we see the value of 
$\theta_b\sim 1.5$ in the DR3 case, which means the $\tb_1$ is dominantly
$\tb_R$. Meanwhile, the red points indicate that
in the HS model, the $\tb_1$ is dominantly $\tb_L$.

\FIGURE[t]{
\includegraphics[width=9cm]{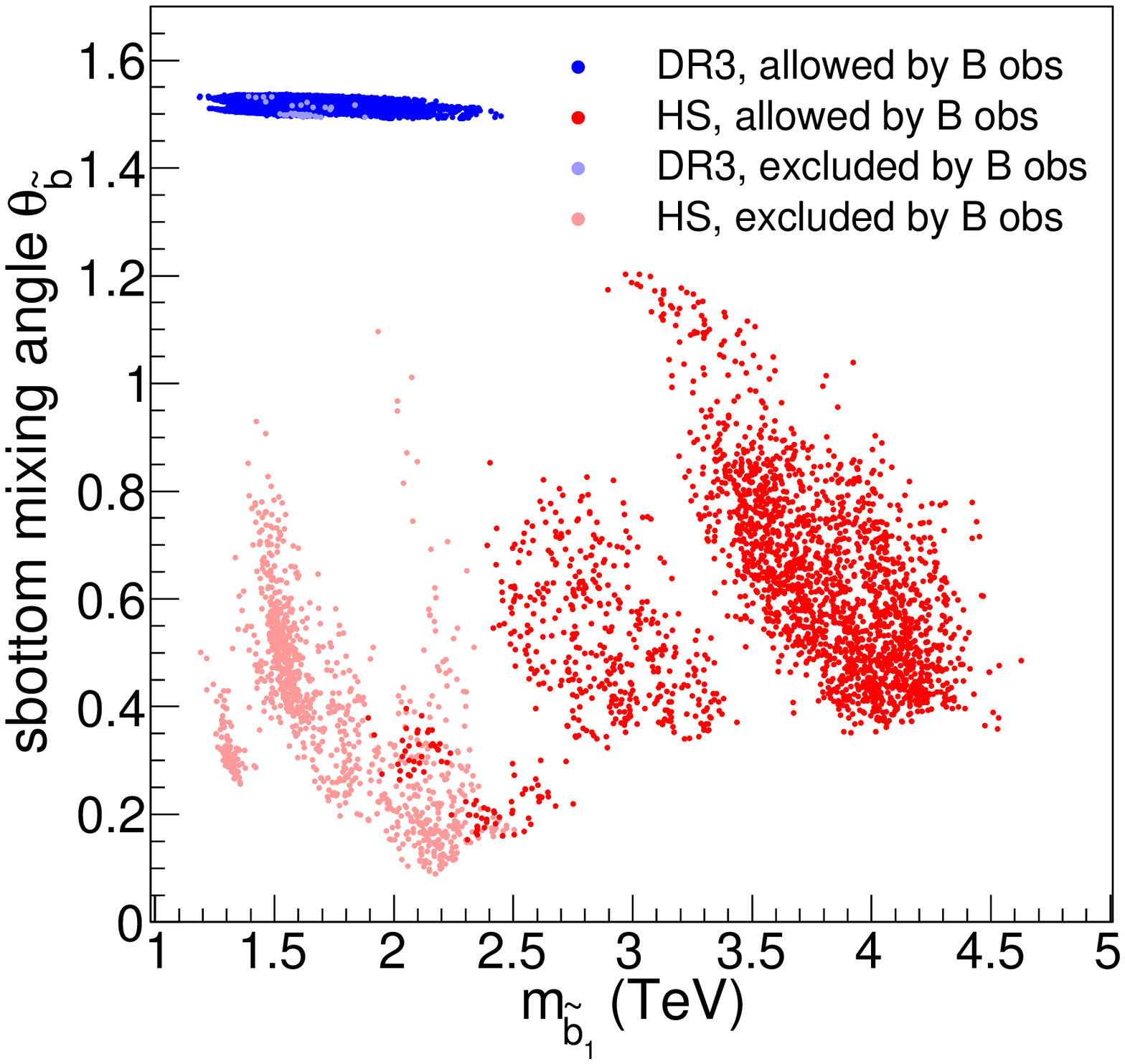}
\caption{Yukawa unified solutions with $R<1.05$ from the ``just-so'' 
HS model (red points) and models with DR3 splitting (blue points) 
in the $m_{\tb_1}\ vs.\ \theta_b$ (rad.) plane; points excluded by $B$-physics 
constraints are shown in lighter colour.
}\label{fig:thb}}

The composition of $\tb_1$ in principle might be measureable at the LHC if
a bottom squark production event sample can be isolated, and the $\tb_1$
branching fractions can be measured. Here note that the $\tb_L$ decays into 
wino-like charginos and neutralinos and/or into $W\tst_1$, while the $\tb_R$ does not. 
To give a concrete example, we compare the $\tb_1$ decays at the DR3 
point of Table \ref{tab:mass} to those of a HS point 
with a very similar $\tb_1$ mass: point A of Ref.~\cite{bhkss}. 
This latter point has $m_{16}=5$~TeV, $m_{\tb_1}=1322$~GeV, 
$m_{\tst_1}=834$~GeV, $m_{\tg}=363$~GeV, $m_{\twpm_1,\tz_2}=109$~GeV 
and $m_{\tz_1}=50$~GeV. 
Several relevant $\tb_1$ branching fractions are listed in 
Table~\ref{tab:BFs}. 
As can be seen, in the DR3 model, $\tb_1$ decays nearly 100\% of the time into
$b\tg$, while in the HS model, there is a sizable branching
into $t\tw^-_1$ and $W\tst_1$ states.

\begin{table}\centering
\begin{tabular}{lcc}
\hline
parameter & Pt.\,B \cite{bhkss}\ & DR3 \\
\hline
$m_{16}(1,2)$& 10000 & 11805.6 \\
$m_{16}(3)$  & 10000 & 10840.1 \\
$m_{10}$   & 12053.5 & 13903.3 \\
$M_D$      & 3287.1 & 1850.6 \\
$m_{1/2}$  & 43.9442 & 27.414 \\
$A_0$      & -19947.3 & -22786.2 \\
$\tan\beta$& 50.398 & 50.002 \\
$R$        & 1.025  & 1.027 \\ 
$\mu$      & 3132.6 & 2183.4  \\
$m_{\tg}$   & 351.2 & 321.4   \\
$m_{\tu_L}$ & 9972.1 & 11914.2   \\
$m_{\tst_1}$& 2756.5 & 2421.6   \\
$m_{\tb_1}$ & 3377.1 & 1359.5  \\
$m_{\te_R}$ & 10094.7 & 11968.5  \\
$m_{\twpm_1}$ & 116.4 & 114.5  \\
$m_{\tz_2}$ & 113.8 & 114.2  \\ 
$m_{\tz_1}$ & 49.2 &  46.5  \\ 
$m_A$       & 1825.9 &  668.3  \\
$m_h$       & 127.8 &  128.6  \\ 
$\theta_b$ (radians)  & 0.329 & 1.53 \\ \hline
\end{tabular}
\caption{Masses in~GeV units and parameters
for  Yukawa-unified point B of Ref.~\cite{bhkss} with
just-so HS, and a point with the DR3 model using $M_N=10^{13}$ GeV.
We also give the $b$-squark mixing angle.
}
\label{tab:mass}
\end{table}

\begin{table}\centering
\begin{tabular}{lcc}
\hline
parameter & Pt.\,A \cite{bhkss}\ & DR3 \\
\hline
$m_{\tb_1}$ & 1321.8 & 1359.5  \\
$\tb_1\to t\tw^-_1$ & 8\% & 0.1\%  \\
$\tb_1\to W\tst_1$ & 30\% & $-$  \\
$\tb_1\to b\tg$ & 55\% & 99\%  \\ 
\hline
\end{tabular}
\caption{Sbottom mass and branching fractions for a HS model 
(point A of Ref.~\cite{bhkss}) and for the DR3 point from 
Table 2. 
Note that the two points have very similar $m_{\tb_1}$, 
but the $\tb_1$ is mainly a 
left-squark in the HS case, while it is mainly a right-squark in the DR3 case.
} 
\label{tab:BFs}
\end{table}

While it is conceivable that the $L-R$ composition of $\tb_1$ might be measured
at LHC, the measurement would likely be very difficult and intricate. However, 
a measurement of the composition of $\tb_1$ would likely be quite  
straightforward at a linear $e^+e^-$ collider with sufficient energy to 
produce $\tb_1\bar{\tb}_1$ pairs. First, the branching fractions of $\tb_1$
would likely be much easier to dis-entangle at an $e^+e^-$ collider than at the
LHC. Second, a linear $e^+e^-$ collider is likely to be constructed with 
polarizable electron beams. The total $e^+e^-\to\tb_1\bar{\tb}_1$ cross
section will be very sensitive to the beam polarization, and the composition
of the $\tb_1$. The beam polarization-dependent sparticle pair production 
cross sections  have been calculated in Ref.~\cite{bmt}, and the results are 
plotted versus the beam polarization $P_L(e^-)$ in Fig.~\ref{fig:sigee}. 
Here, we see that
$\sigma (e^+e^-\to \tb_1\bar{\tb}_1)$ peaks at $P_L(e^-)=+1$ in the case of
the HS model, whereas just the opposite peak at $P_L(e^-)=-1$ is expected
in the DR3 case.

\FIGURE[t]{
\includegraphics[width=9cm]{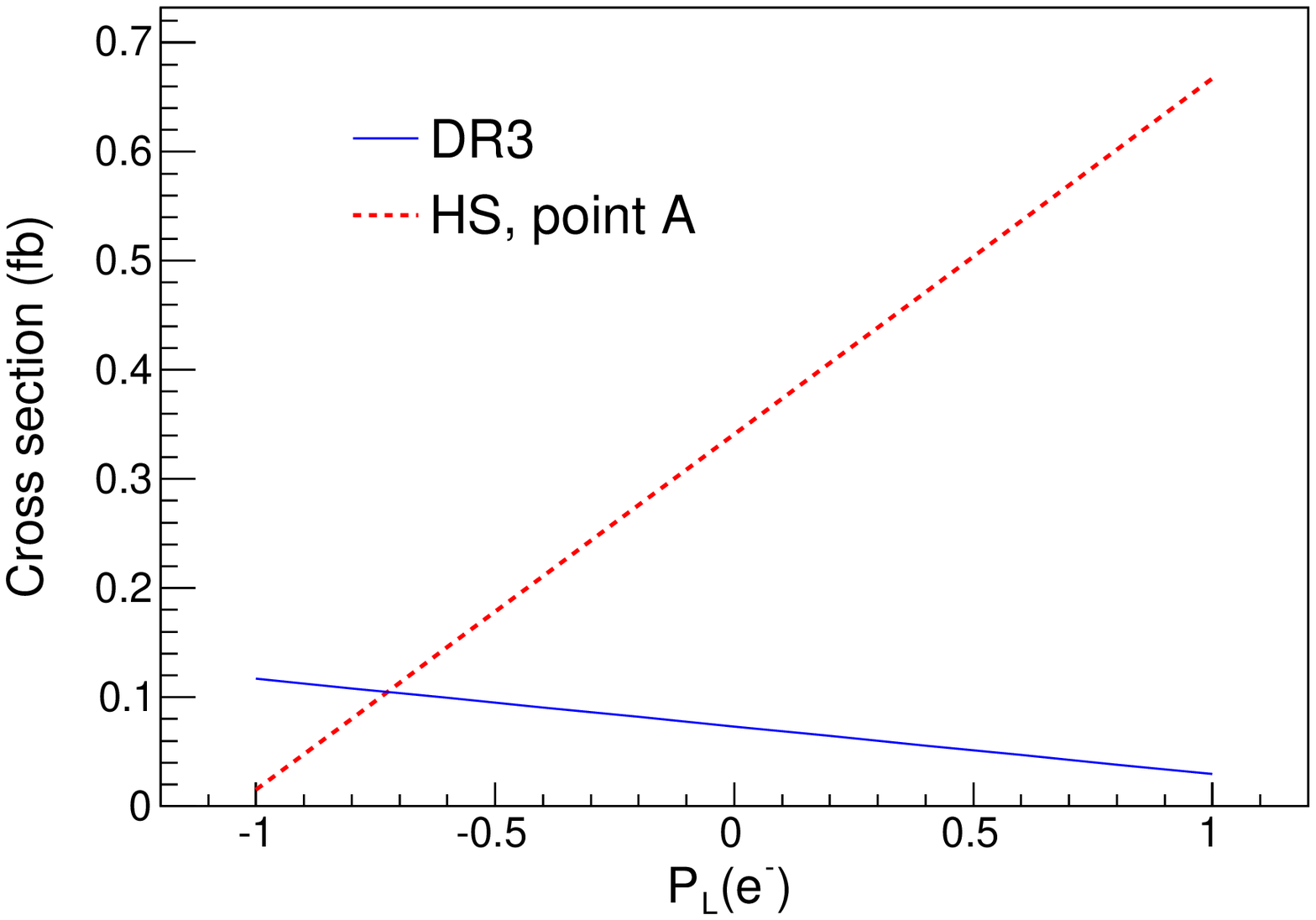}
\caption{Sbottom production cross section, $\sigma (e^+e^-\to \tb_1\bar{\tb}_1 )$, 
in fb versus electron beam polarization $P_L(e^- )$ for
the CLIC accelerator with $\sqrt{s}=3$ TeV.
The blue curve corresponds to a DR3 model point while the red curve 
corresponds to the HS model point A of Ref.~\cite{bhkss}.
}\label{fig:sigee}}

In Fig.~\ref{fig:mA}, we show the value of $m_A\ vs.\ m_{16}$ for
Yukawa-unified models with $R<1.05$ in the HS and DR3 cases.
For moderate to large $\tan\beta$, at tree level we roughly expect
that $m_A^2\sim m_{H_d}^2-m_{H_u}^2$, {\it i.e.} that the value
of $m_A^2$ is nearly equal to the weak scale mass splitting between the
two Higgs soft masses. In the DR3 model, where a much smaller Higgs splitting
is needed at the GUT scale in order to accomplish REWSB, we also find a 
significantly smaller value of $m_A$ expected for a given value of $m_{16}$.
For $m_{16}\sim 10$ TeV, we typically get $m_A\sim 1$ TeV in the DR3 model, while
$m_A\sim 3$ TeV in the HS model. 

The value of $m_A$ may be readily established at the LHC, especially for
the large $\tan\beta\sim 50$ case expected for Yukawa-unified models.
In the large $\tan\beta$ case, the $b$-quark Yukawa coupling is large, and
production cross sections for $A$ are enhanced, both via glue-glue fusion
(triangle diagrams) and via $bA$ and $b\bar{b}A$ production. 
Then the $A$ is typically expected to decay into modes such
as $b\bar{b}$, $\tau^+\tau^-$ and $\mu^+\mu^-$. The first two of these offer
a rough mass bump with which to reconstruct $m_A$; the latter mode into
$\mu^+\mu^-$ suffers from a small branching fraction, but 
may offer a sharper mass bump since all the decay products are
easily detected\cite{chung}. 
However, for extremely high energy muons, the momentum resolution---
determined by track bending in the detector magnetic field--- gets more difficult
at high energies. Evaluation of the $A$ mass bump in all three of these modes may
allow good resolution on the $A$ mass reconstruction. If a measurement of 
$A\to\mu^+\mu^-$ is possible, then it may also be possible to extract information
on the width $\Gamma_A$, which is very sensitive to the value of $\tan\beta$.

\FIGURE[t]{
\includegraphics[width=9cm]{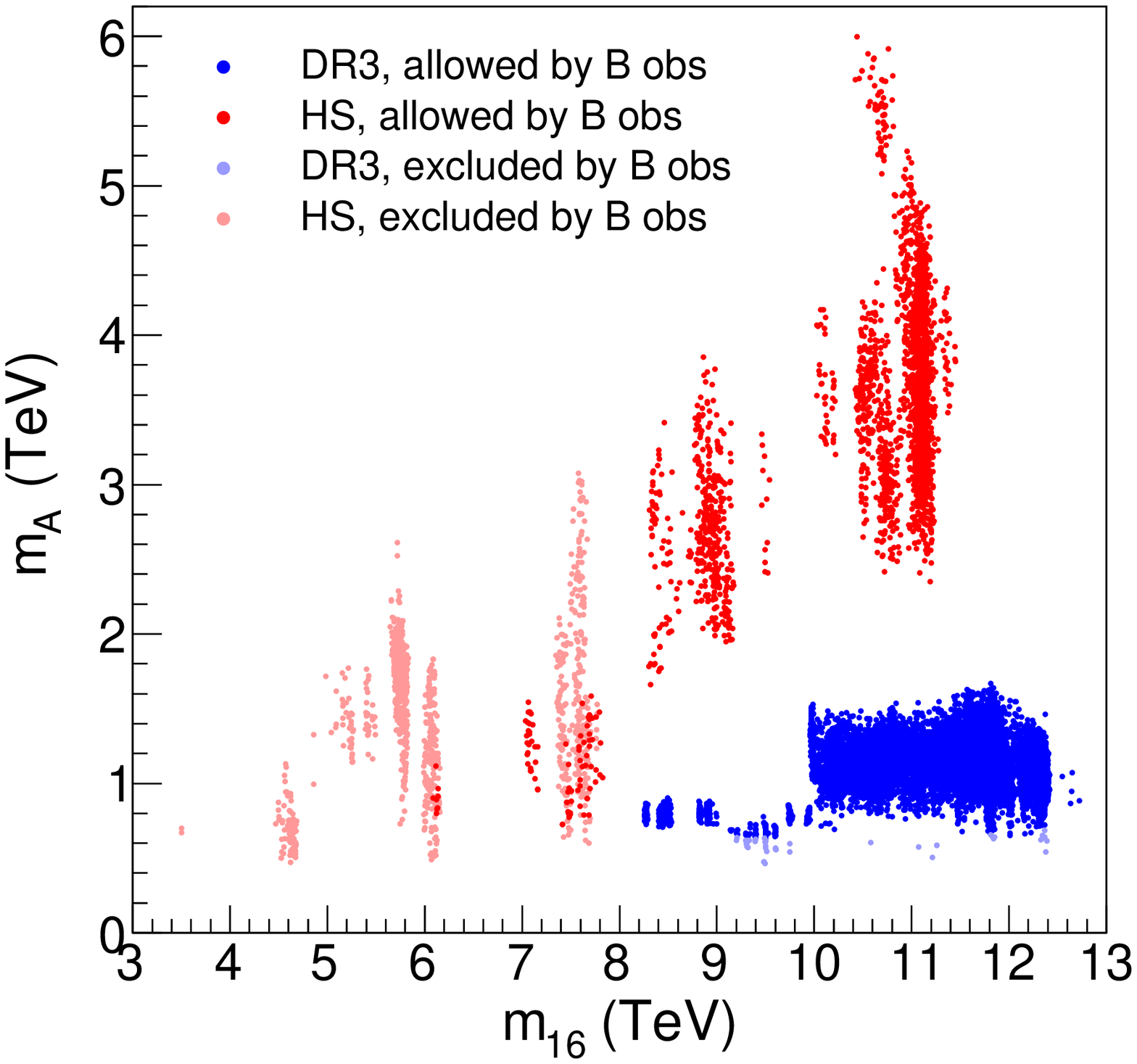}
\caption{Yukawa unified solutions with $R<1.05$ from ``just-so'' 
HS model (red points) and models with DR3 splitting (blue points) 
in the $m_{16}\ vs.\ m_A$ plane; points excluded by $B$-physics 
constraints are shown in lighter colour.
}\label{fig:mA}}

\subsection{Predictions for $B$-physics observables}
\label{ssec:bsg}

For completeness, we present in Fig.~\ref{fig:bsg} the explicit results for the 
branching ratios of the $b\to s\gamma $ and $B_s\to\mu^+\mu^-$ decays
for the Yukawa-unified solutions with $R<1.05$ of the previous section. 

For the HS model, we see that most of the solutions with $m_{16}\alt 8$ TeV lead to
too low a value of BR$(b\to s\gamma )$, and hence are excluded. Of the remaining  
HS points with $m_{16}\alt 8$ TeV, a large fraction has too high a 
BR($B_s\to\mu^+\mu^-$). 
In the end, only a few points with $m_{16}\alt 8$ TeV survive and they have 
BR$(B_s\to\mu^+\mu^-)={\cal O}(10^{-8})$. The HS points with  $m_{16}> 8$ TeV 
mostly comply with the $B$-physics constraints and have 
BR$(B_s\to\mu^+\mu^-)={\cal O}(10^{-9})$.

The situation is different for the DR3 model points, for which $m_{16}$ is always greater 
than 8~TeV. These points cluster around BR$(b\to s\gamma )\sim 3\times 10^{-4}$, 
with only a very small fraction (less than 1\%) excluded by a too high 
BR($B_s\to\mu^+\mu^-$). Indeed the excluded points are those with the lowest 
$m_A$ values, cf.\ Fig.~\ref{fig:mA}. Interestingly, owing to the lower $m_A$, 
the DR3 model predicts BR$(B_s\to\mu^+\mu^-)={\cal O}(10^{-8})$, which may 
well be probed in the near future at the Fermilab Tevatron collider. 

\FIGURE[t]{
\epsfig{file=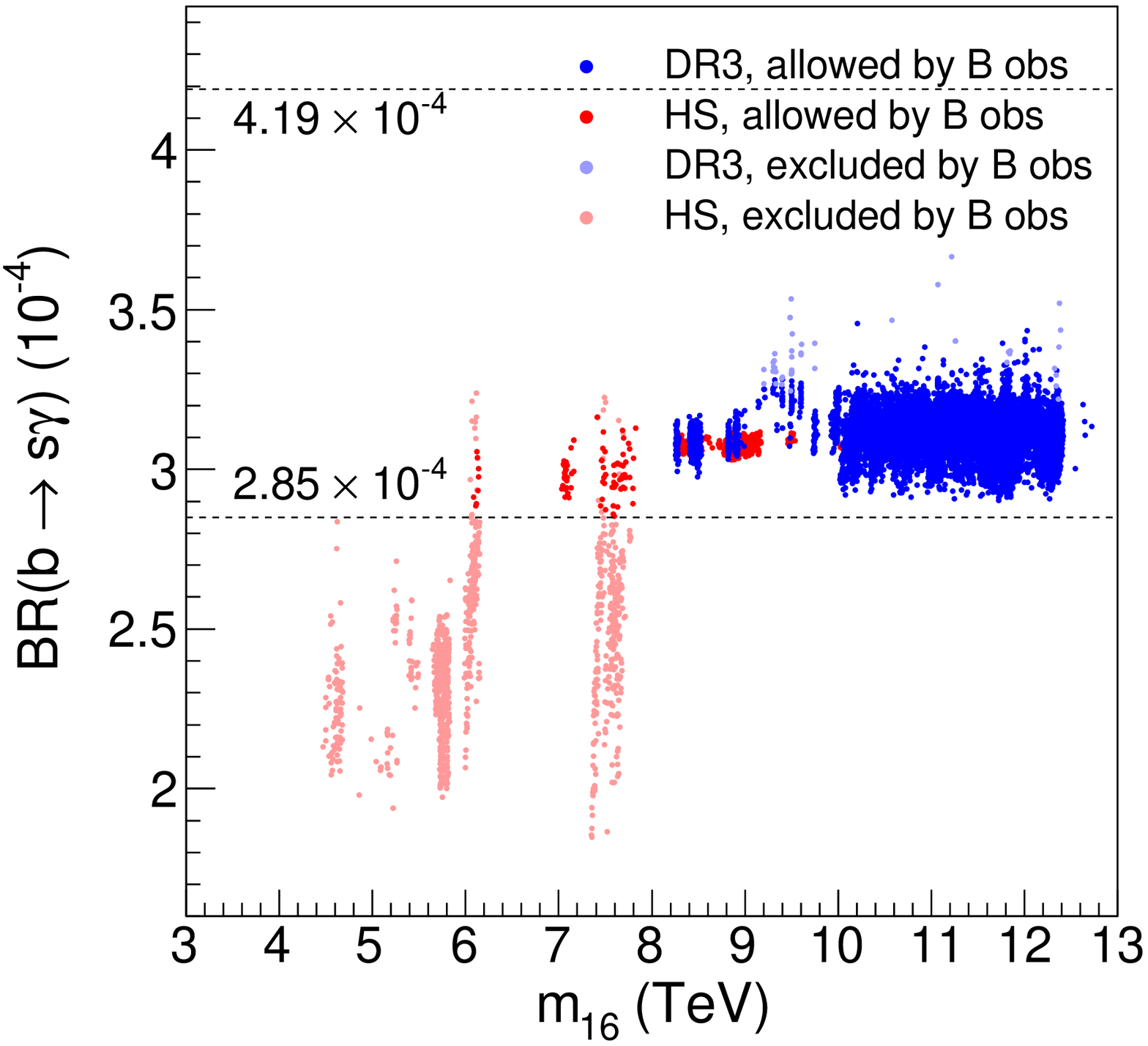,width=7cm,angle=0}
\epsfig{file=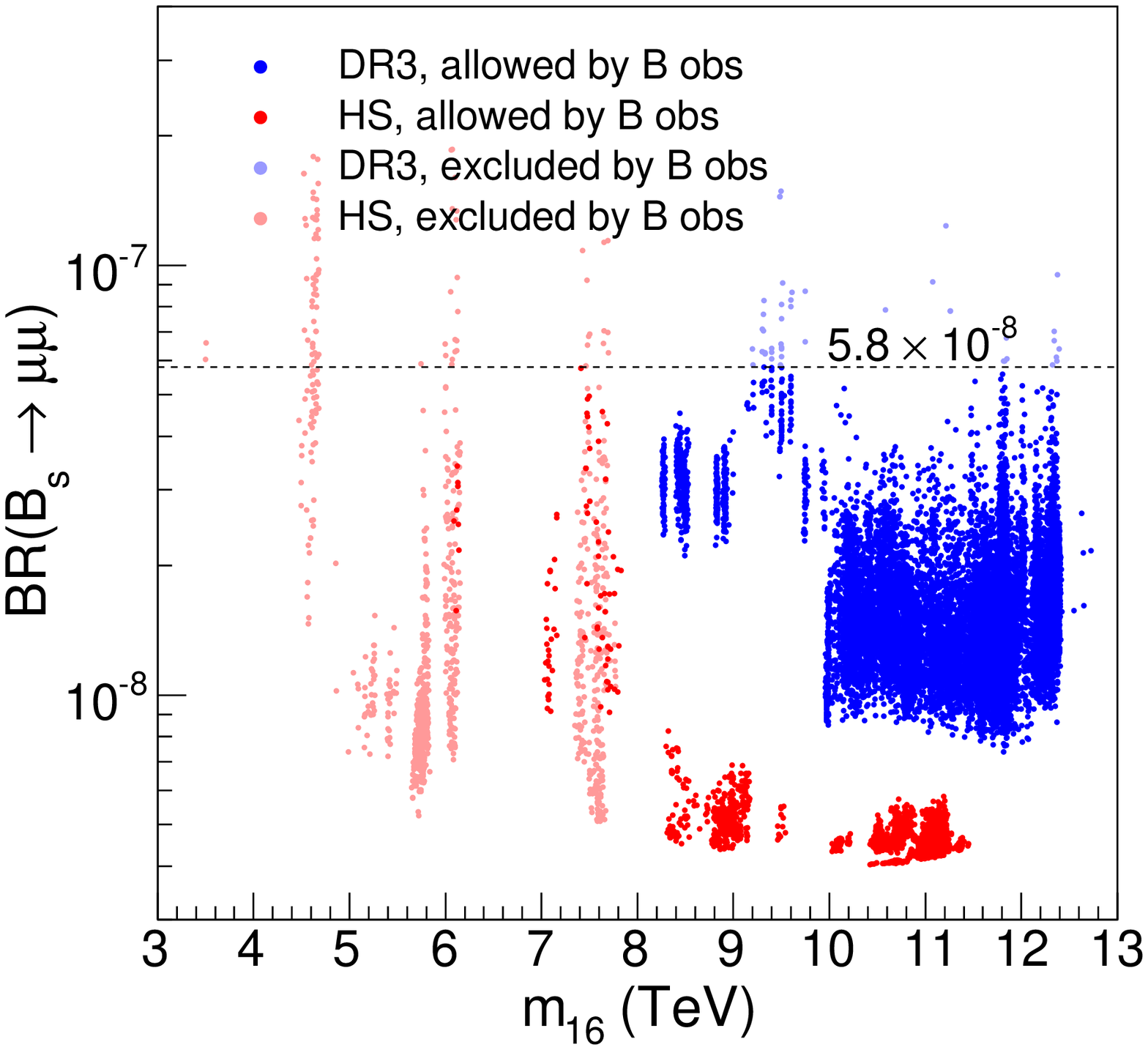,width=7cm,angle=0}
\caption{\label{fig:bsg} Predictions for BR$(b\to s\gamma) \ vs.\ m_{16}$ (left plot) 
and BR$(B_s\to\mu^+\mu^-) \ vs.\ m_{16}$ (right plot) from Yukawa-unified 
models in both the HS (red dots) and DR3 (blue dots) models. 
Also shown are the $2\sigma$ limits used in our analysis.}}

\section{Summary and conclusions}
\label{sec:conclude}

In this paper, we have re-investigated the issue of electroweak symmetry breaking 
in Yukawa-unified SUSY models. It is well known that perfect Yukawa coupling unification 
can be achieved in simple $SO(10)$ SUSY GUTs 
with universal soft-breaking terms for gauginos and sfermions, but non-universal 
Higgs-mass terms. 
However, it is hard to understand how this ``just-so'' Higgs splitting may arise only
in the Higgs multiplets, and not in the corrsponding matter multiplets,
as is expected in $D$-term splitting. We have found that Yukawa coupling 
unification good to few per cent can be achieved in the 
case of $D$-term splitting but only if one also allows for the presence of
the neutrino Yukawa coupling, along with first/second versus third
generation matter scalar splitting. Each of these three features---
$D$-term splitting, right hand neutrino Yukawa couplings and 
third generation splitting, together comprising the DR3 model--- 
are to be expected in simple GUT models based on $SO(10)$. 
In this case, the DR3 model may be considered
more satisfying from the $SO(10)$ GUT point of view than the HS model.

The two models lead to many similarities in the expected sparticle mass spectra,
but also some important differences. Regarding similarities, both lead to
a split spectra with first/second generation scalars in the $\sim 10$ TeV regime,
with third generation and heavy Higgs scalars in the few TeV regime, along
with a very light spectrum of gauginos. In particular, with gluinos expected
in the $300-500$ GeV mass range, a robust variety of gluino pair production
events are expected at the CERN LHC\cite{so10lhc}. Regarding differences, 
the amount of Higgs splitting for a given value of $m_{16}$ is expected to be much
less in the DR3 model, leading to much lighter values of $m_A$, $m_H$ and
$m_{H^\pm}$. These heavier Higgs states stand a much higher chance to be detectable
at LHC in the DR3 model, as compard to the HS model. In addition, the lightest
bottom squark is expected to be much lighter for a given value of
$m_{16}$ in the DR3 model, as compared to the HS model. Also, the smoking gun 
difference is that the $\tb_1$ should
be predominantly a right-squark in DR3, while it is expected to be dominantly 
a left-squark in the HS model. This can in principle be detected at LHC due to
the different $\tb_1$ branching fractions which are expected; however, in practise, this
differentiation is likely to be a difficult enterprise. It will be much simpler
at a CLIC-type $e^+e^-$ linear collider operating with $\sqrt{s}>2m_{\tb_1}$. In this case,
the $\tb_1$ production and decay modes should be more readily identified, and especially the
total $\tb_1\bar{\tb}_1$ production cross section will depend on electron beam polarization 
in very different fashions for the two models. In this case, the two models should
then be easily distinguishable.

We also evaluated predictions for BR$(b\to s\gamma )$ and
BR$(B_s\to\mu^+\mu^- )$ in the HS and DR3 models. In the HS model, 
most points with $m_{16}<8$ TeV are excluded by these constraints. In the case
of the DR3 model, which apparently requires $m_{16}\agt 8$ TeV, $99\%$ of the 
points are
allowed by the $B$-decay constraints. But the predicted rate for the $B_s\to\mu^+\mu^- $
decay is just below its current experimental limit, and may well be probed in the near future
as data continues to accrue at the Fermilab Tevatron collider. 

Finally we note that a large RHN Yukawa coupling, as assumed in this study, 
can lead to lepton flavor violation (LFV) by generating off-diagonal LFV terms 
in the charged-slepton mass matrix through RG running 
\cite{Borzumati:1986qx,Dedes:2007ef}.   
This can lead to additional constraints which may further sharpen the predictions 
for Yukawa-unified models. This issue is left for future work \cite{future}. 

{\it Note added:} As this paper was being finalized, some related papers appeared on Yukawa coupling unification in models without universality in gaugino masses\cite{schafi} and $A$-terms\cite{guad}.

\acknowledgments
The work of SK is supported in part by the French ANR project ToolsDMColl, 
BLAN07-2-194882. The work of SS is supported in part by the US Department of
Energy grant number DE-FG-97ER41022.

%

\end{document}